# Tuning a Circular p-n Junction in Graphene from Quantum Confinement to Optical Guiding

Yuhang Jiang[1†], Jinhai Mao[1†], Dean Moldovan[2†], Massoud Ramezani Masir [2, 3], Guohong Li[1],
Kenji Watanabe[4], Takashi Taniguchi[4], Francois M. Peeters[2] and Eva Y. Andrei[1]

[1]Department of Physics and Astronomy, Rutgers University, 136 Frelinghuysen Road, Piscataway, NJ 08855 USA

[2] Departement Fysica, Universiteit Antwerpen, Groenenborgerlaan 171, B-2020 Antwerpen, Belgium

[3]Department of Physics, University of Texas at Austin, Austin TX 78712, USA

[4]Advanced Materials Laboratory, National Institute for Materials Science, 1-1 Namiki, Tsukuba 305-0044, Japan

† These authors contributed equally to this work.

**The motion of massless Dirac-electrons in graphene mimics the propagation of photons[1,2]. This makes it possible to control the charge-carriers with components based on geometrical-optics and has led to proposals for an all-graphene electron-optics platform[3-9]. An open question arising from the possibility of reducing the component-size to the nanometer-scale is how to access and understand the transition from optical-transport to quantum-confinement. Here we report on the realization of a circular p-n junction that can be continuously tuned from the nanometer-scale, where quantum effects are dominant, to the micrometer scale where optical-guiding takes over. We find that in the nanometer-scale junction electrons are trapped in states that resemble atomic-collapse at a supercritical charge[10-14]. As the junction-size increases, the transition to optical-guiding is signaled by the emergence of whispering-gallery modes**[15-17] **and Fabry-Perot interference**[6,18]**. The creation of tunable junctions that straddle the crossover between quantum-confinement and optical-guiding, paves the way to novel design-architectures for controlling electronic transport.**

The photon-like propagation of the Dirac-electrons in graphene together with the record-high electronic mobility afforded by Klein-tunneling[19] can lead to applications based on ultra-fast electronic response and low dissipation[20,21]. But, while Klein-tunneling is responsible for the high mobility resulting from the suppression of backscattering, it also makes it difficult to control or shut off carrier-motion, to the detriment of electronic switching. We show that it is possible to control the charge-carriers with a circular p-n junction consisting of a central region acting as a circular potential barrier[22,23]. The junction is created by combining a planar back-gate with a top-gate consisting of an Au decorated tip and it can be continuously tuned from the nanometer to the micrometer scale. The nanometer junction is defined by a deep and narrow potential-well created by the tip-induced charge. It traps the Dirac-electrons in quantum-confined states which are the graphene equivalent of the atomic collapse states (ACS) predicted to occur at super-critically charged nuclei[10,11]. As the junction size increases, the transition to the optical regime is marked by the appearance of whispering-gallery modes (WGM)[17], which are the graphene equivalent of the well-known WGMs observed at the perimeter of acoustic[15] or optical resonators[16] and by the appearance of a Fabry-Perot interference pattern for junctions close to a boundary.

In the dual-gate device employed here, the back-gate electrode tunes the global carrier-density, and the top-gate controls the local doping under the STM tip (Fig. 1A). The variable-size p-n junction is realized by controlling the back gate voltage, $V_{bg}$ (Fig.1B). Prior to creating the p-n junction we measure the gate dependence of the tunneling resistance (dI/dV) spectra[24,25] with a pristine Pt-Ir tip, which for low bias-voltages ($V_b$) is known to be noninvasive[25]. By contrast, a strongly biased STM tip can create a sufficiently deep potential-well that traps electrons in quantum-confined states[11,26]. To access this regime we functionalize the STM tip by gently poking it into an Au electrode to create a large tip-graphene work-function difference, ΔW. The resulting tip-induced potential depends the tip-shape, which can be controlled by the speed of its retraction from the Au electrode. Slow retraction rates (0.1 nm/s) decorate the tip with a sharp protruding Au nanowire[27,28], while rapid retraction coats it with an Au film (see methods and SI for details). The effectiveness of the procedure is assessed by comparing the tunneling spectra obtained before and after the functionalization. In Fig 2 we focus on the spectra obtained with the sharp nanowire decorated tip (see SI for spectra obtained with Au coated tips, and tip calibration). Prior to tip decoration (Fig.2A) the spectra exhibit the well-known "V" shape of unperturbed graphene[25] with a minimum that shifts according to the gate induced doping. In contrast, the spectra taken with the Au nanowire tip are qualitatively different and display a strong gate dependence. In the p-doped regime (Fig. 2B) they feature a distinct negative energy resonance peak below the bulk Dirac-point (DP) energy ($E_D$). This peak is very sensitive to the global doping level: in the highly p-doped regime ($V_g = -54$V) it is a barely distinguishable broad feature near the bulk $E_D$ which becomes more pronounced and shifts monotonically down in energy as doping is decreased. As p-doping is further reduced a second peak materializes at $V_g = -6$V.

The spectral features in Fig. 2B resemble those of ACS at super-critically charged impurities in graphene[12,14]. The ACS are negative energy states that can bind Dirac-electrons to a local charge, Z. They emerge when the interaction strength, as quantified by the Coulomb-coupling constant, $\beta \equiv Z\alpha_g$, exceeds a critical value $\beta_c \geq 0.5$ .Here

$\alpha_g = \alpha \frac{c}{\kappa v_F}$, α is the QED fine-structure constant, $c$ the speed of light, $v_F$ the Fermi-velocity, and κ the effective dielectric-constant. Similar to the negative energy resonances observed here, ACS at supercritically charged vacancies are observed in the p-doped regime only, their energy becomes more negative with reduced carrier density, and for sufficiently large charge and low doping a second resonance appears just below $E_D$. These similarities suggest that the resonances observed here reflect ACSs associated with the tip-induced local charge[11,26].

To gain insight into these results we used numerical tight-binding to calculate the local DOS (LDOS) in the presence of the sharp tip, which was modeled as a conical segment that broadens into a spherical shell farther away (Fig. 6s). Using Thomas-Fermi theory to account for screening we calculated the tip-induced potential in the graphene layer as a function of $V_b$, $V_g$ and $\Delta W$ (S4, S7). The tip-induced potential for a tip with $\Delta W = 0.7$V is shown in Fig. 1B. With this potential as input, we calculated the gate-voltage dependence of the LDOS (Fig. 2C, 2D). The calculated spectra closely match the measured ones: in the highly p-doped regime ($V_g = -54$V), there is a broad feature near the bulk $E_D$; as the carrier concentration decreases ( $-19$ V $< V_g <$ $-6$V) a sharp peak appears below $E_D$ that dives into the negative-energy sector, and finally a second peak materializes just below $E_D$. Significantly, the calculated resonance peak energy strongly depends on $\Delta W$ and disappears below a critical-value $\Delta W_c \sim 0.3$eV as shown in Fig.2E. This indicates that, similar to the ACS case, there exists a control parameter, $\Delta W$, which has to exceed a critical-value for the negative-energy resonances to appear. This may not be surprising considering that the sharp tip with its large $\Delta W$ creates a deep and narrow potential which resembles that of a point-charge (Fig. 1B). To make the connection with charge-induced ACS we define an effective tip-induced point-charge which, over a circular area of radius R, induces the same amount of charge as that of the tip-induced potential (SI). This effective point-charge results in an effective Coulomb-coupling constant:

$$\beta_{\mathit{eff}} = \frac{1}{\hbar v_F}\sqrt{2\int_{r_0}^{R} V(r)^2 r\, dr /(1+2\ln(R/r_0))}\,,$$ where $r_0$ ~0.25nm is a lower cutoff at the scale of a graphene unit cell and R is taken at the crossing between the steep and shallow parts of the potential.

The gate dependence of the calculated $\beta_{\mathit{eff}}$ for several values of $\Delta W$ is shown in Fig. 2F. Comparing to the experimental data in Fig. 2B we find that the $\beta_{\mathit{eff}}$ ($V_g$) curve with $\Delta W = 0.7$eV reproduces the observed features: similar to the data, this curve attains its maximum value for $V_g$ ~ -6V corresponding to the largest attainable shift of the ACS peak relative to the DP. To determine the critical value, $\beta_{\mathit{eff}}^c$, we note that the resonance peak disappears when it crosses the DP at $V_g$ ~ -30V (Fig. 2D) and that $\beta_{\mathit{eff}}$ attains super-critical values only for $\Delta W \geq 0.3$V. Both these constraints are satisfied by choosing $\beta_{\mathit{eff}}^c$ = 0.7 (dashed-line in Fig. 2F). This value is slightly higher than that expected for a pure Coulomb potential, 0.5, which is not surprising given that the tip-induced charge is more spread out than that of a point charge. Furthermore, we find that similar to the experimental data, $\beta_{\mathit{eff}}$ increases with reduced carrier density in the p-doped regime and, after peaking, it sharply drops below the critical value upon approaching the DP. These features suggest that the tip-induced negative energy peaks observed here, similarly to resonances induced by a point-charge [11,12,14], can be attributed to atomic collapse physics.

Next, we study the evolution of the LDOS with junction size (Fig. 3). For $V_g$ = -14V, corresponding to the ACS regime discussed above, the nanometer sized junction traps electrons in a quasi-localized state at its center (Fig.3A, E). The crossover to the micron-size junction is controlled by tuning $V_g$ to gradually approach the bulk DP. This leads to an increase in screening length which broadens the top part of the potential well (Fig.3B, C). The induced charge spreads out into an extended pool residing in a circular cavity whose boundaries are defined by the sharp kink forming the circular p-n junction at the crossing of the potential with the Fermi energy (dashed line in Fig. 7s(a)). The cavity can form within a narrow range of doping below the DP. Its radius, $R_c$, grows with increasing $V_g$ and diverges at the bulk charge-neutrality point, beyond which it disappears. In graphene the reflection of electrons at a *p-n* junction is governed by Klein scattering which, for certain oblique angles, can produce perfect reflection[19,23]. As a result, the cavity can support WGMs when the conditions of total internal reflection and constructive interference are met. Interestingly, for off-resonance electron energies the junction serves as a lens with tunable focal-length that can guide electronic motion[22].

With this understanding in mind we examine the STS spectra close to the charge neutrality point (Fig.3G). Here we observe a succession of nearly equally spaced peaks (red arrows) below the bulk $E_D$ which are consistent with WGMs within the cavity defined by the broad circular p-n junction, see Fig.5s for the details on the $V_g$ dependence. Even though the WGMs observed with the sharp tip are seen only in a narrow doping window they share the same characteristics as those formed by a blunt Au coated tip as shown in Fig.4s, and Fig.5s, consistent with earlier reports[17]. Tuning the gate voltage into the n-doped regime results in an n+-n junction (Fig. 3D, 3H). In this case, with the outer p-doped regime missing, the boundary that sustained WGMs no longer exists and the LDOS becomes featureless. Notably, even though the potential profile has reverted back to the point-charge-like potential, the ACS peak is absent. This can be understood from the doping dependence of $\beta_{eff}$ (Fig. 2F) which sharply dives below the critical value at the crossing into the n-doped regime[14]. In this regime the circular junction acts as a weak scatterer[22]. The different regimes of the p-n junction are elegantly illustrated by the $\beta$ $\log E_n \propto n_{eff}(V_g)$ curve in Fig. 2F: -30V < $V_g$ < -6V defines the nano-scale junction regime where ACSs are observed; for -6V < $V_g$ ≤ 0V the junction crosses over into the micron-scale regime where WGMs emerge. Finally, the sequence of peak energies does not follow the characteristic $\log E_n \propto n$ of ACS resonances[11,29], excluding their interpretation in terms of high order ACSs.

In Fig. 4 we take a closer look at the crossover regime. Close to charge neutrality where the cavity has almost disappeared the LDOS still shows discrete negative-energy modes, $V_b$ < 0, corresponding to WGM, but the positive-energy sector, $V_b$ > 0, is featureless (Fig 3G). The absence of structure for $V_b$ > 0 indicates that although the electrons in this regime are weakly scattered by the tip-induced potential[22], the LDOS is unaffected. This situation is changed by placing a reflecting boundary close to the junction. The interference of the electron beams trapped between the tip-potential and the boundary produces a Fabry-Perot pattern, which is strongly enhanced at the tip position due to the cavity focusing effect (Fig. 9s). This produces a pronounced sequence of equally-spaced positive-energy peaks in the LDOS, $\Delta E = \pi \hbar v_F / d$, where d is the tip-boundary distance, as shown by the simulation in Fig. 4A. The data in Fig 4B reveals the effect of a boundary defined by an Au electrode (Fig 4C) placed at a distance $d \sim 100$nm from the junction. We note 6 equally-spaced ($\Delta E \sim 19$meV) positive-energy peaks (above the DP) which are consistent with the generation of Fabry-Perot interference resulting from reflections off the electrode (SI). The *1/d* position dependence of the measured peak energy separation is in good agreement with numerical simulations of Fabry-Perot interference peaks, as shown in Fig. 4C inset. Further support for this picture is provided by the disappearance of the positive energy peaks for distances larger than d ~ 250nm, from which we estimate a phase coherence length of ~ 500nm in this sample.

The circular p-n junctions discussed here open novel nano-electronic design opportunities. Devices incorporating graphene nano-junction arrays consisting of nanowires with large $\Delta W$ such as Ti, Cu, or Ta deposited on gated graphene could enable both switching and guiding of Dirac-electrons, could add an important building block to the electron-optics toolbox.

### Methods

Samples consist of twisted double layer exfoliated graphene which are deposited on an hBN flake by a dry transfer process. The hBN flake is exfoliated on 300nm $SiO_2$ dielectric layer capping a highly doped Si substrate serving as a back-gate. After annealing in forming gas at 300 °C for 3 hrs, the first graphene layer is transferred onto the hBN with a PMMA/PVA sacrificial film which is subsequently removed by acetone. The second layer is deposited following the same steps. Subsequently Au/Ti electrodes are deposited using standard SEM lithography followed by overnight annealing. STM measurements are performed in a home-built instrument at 4.2K using a cut $Pt_{0.8}Ir_{0.2}$ tip[30,31] which is known to be non-invasive within the typical range of experimental bias voltages, $V_b < 0.2$V. The dI/dV spectra are collected by the standard lock-in method with a 2mV, 473.1Hz modulation added to the DC bias. We define the gate voltage $V_g = V_{bg} - V_0$, as the applied backgate voltage, $V_{bg}$, measured with respect to the backgate voltage at charge neutrality, $V_0$. Details of the tip functionalization are described in the SI.

Funding provided by DOE-FG02-99ER45742 (STM/STS), NSF DMR 1708158 (fabrication). Theoretical work supported by ESF-EUROCORES- EuroGRAPHENE, FWO-VI and Methusalem program of the Flemish government.

**Figure Captions:**

**Fig.1 Tunable circular p-n junction.** (**A**) Variable size graphene junctions are produced with a dual gate configuration consisting of a silicon-substrate back-gate ($V_{bg}$) that controls the global doping level and an STM tip serving as a top gate ($V_{tg}$) to tune the local carrier concentration underneath the tip. Top-left: nanometer scale graphene p-n junction created by gating the bulk in the deep p-doped regime. Bottom-left: Wavefunction distribution for a WGM with angular momentum m=10 supported by a 300nm junction created in the weakly p-doped regime. Right panel: back-gate dependence of the tip-induced potential showing the Coulomb-like potential giving rise the ACS (top) and the cavity-like potential responsible for the WGM (bottom). The dashed line indicates the position Fermi level. (**B**) Calculated potential-profile for a series of back-gate voltages and a tip-graphene work-function difference $\Delta W$ =0.7eV.

**Fig.2 Gate voltage dependence of the LDOS in a nano-scale graphene p-n junction.** (**A**) Back-gate dependence of dI/dV curves before tip functionalization. In all the panels the arrows label the bulk DP. (**B**) Same as (**A**) after the tip functionalization with an Au nanowire. ( STM setpoint: $V_b$ = -50mV, $I$ = 0.1nA). (**C**) Calculated LDOS for the graphene nanojunction for $\Delta W$ = 0.7eV. (**D**) 2D map of the LDOS evolution with $V_g$. Black dashed curve indicates the bulk DP. Red dashed lines correspond to the curves in (**C**). (**E**) 2D map of the LDOS evolution with $\Delta W$. The dashed line represents the bulk DP for $V_g$. = -9V. (**F**) Evolution of $\beta_{eff}$ with $V_g$ for several values of $\Delta W$ indicated in the legend.

**Fig.3 LDOS in the variable size p-n junction.** (**A to D**) Simulated map of the LDOS as a function of position relative to the center of the tip for indicated back-gate voltages. The dotted lines indicate the potential profile. (**E to H**) Measured dI/dV curves for different

back-gate voltages corresponding to the simulated maps in A to D. The arrows in panel G indicate the WGM peaks and the red curve correspond to the same data after background subtraction. Insets illustrate the p-n junction boundaries. Red: highly n-doped regime; yellow: n-doped regime; blue: p-doped regime.

**Fig.4 Fabry-Perot interference-pattern.** (**A**) Calculated LDOS for the graphene junction showing WGMs and Fabry-Perot resonances due to reflections from a nearby boundary. $E_D$ labels the bulk Dirac point. (B) Measured dI/dV spectrum at a tip-induced p-n junction for $V_g$ =-0.6V recorded at the position marked by the square in (C). The black arrow labels the bulk Dirac point. (STM setpoint: $V_b$ = -50mV, $I$ = 0.1nA, $V_g$ = -0.6V). (C) STM topography of the graphene sample close to the Au electrode boundary (dashed line). Inset: evolution of peak period with the distance to the boundary as measured (symbols) and theoretical curve (black line) $\Delta E = \pi \hbar v_F / d$.

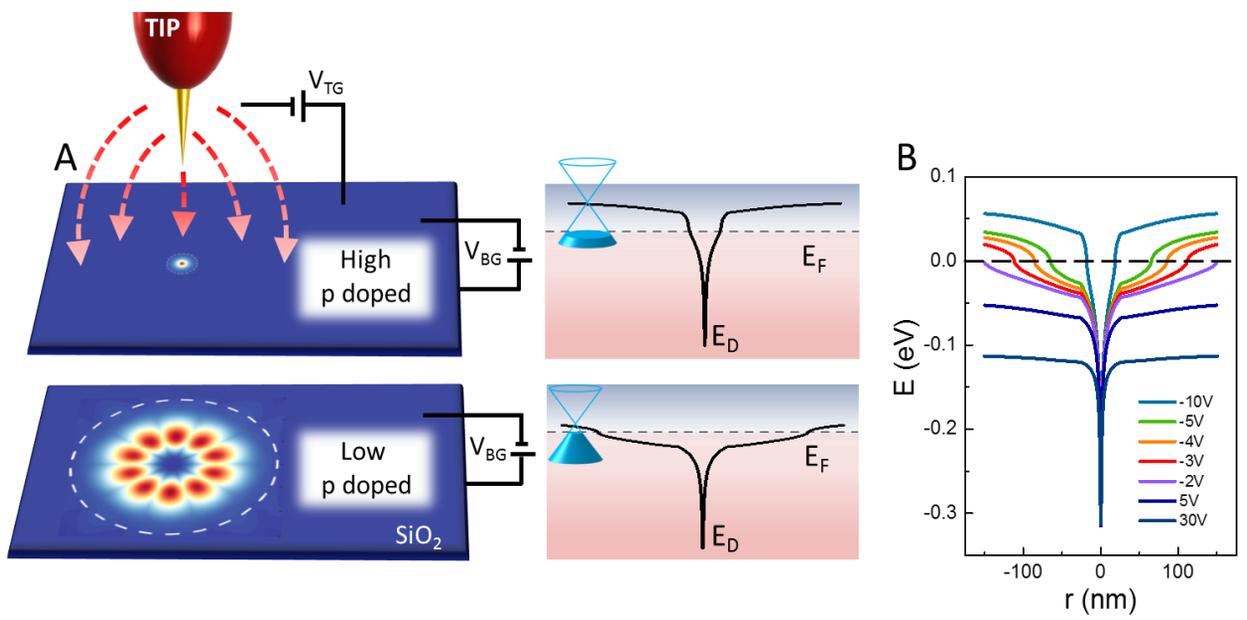

**Figure 1**

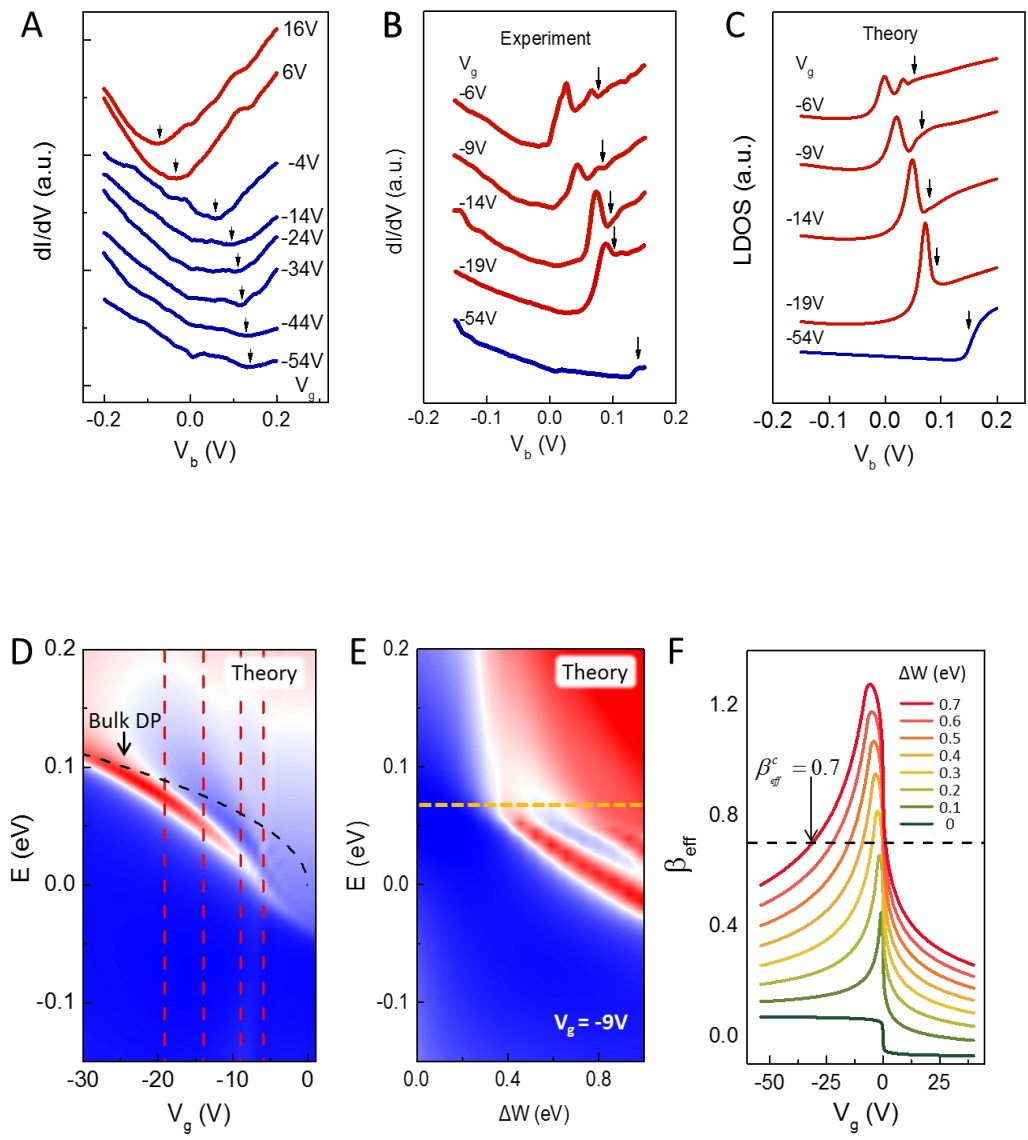

**Figure 2**

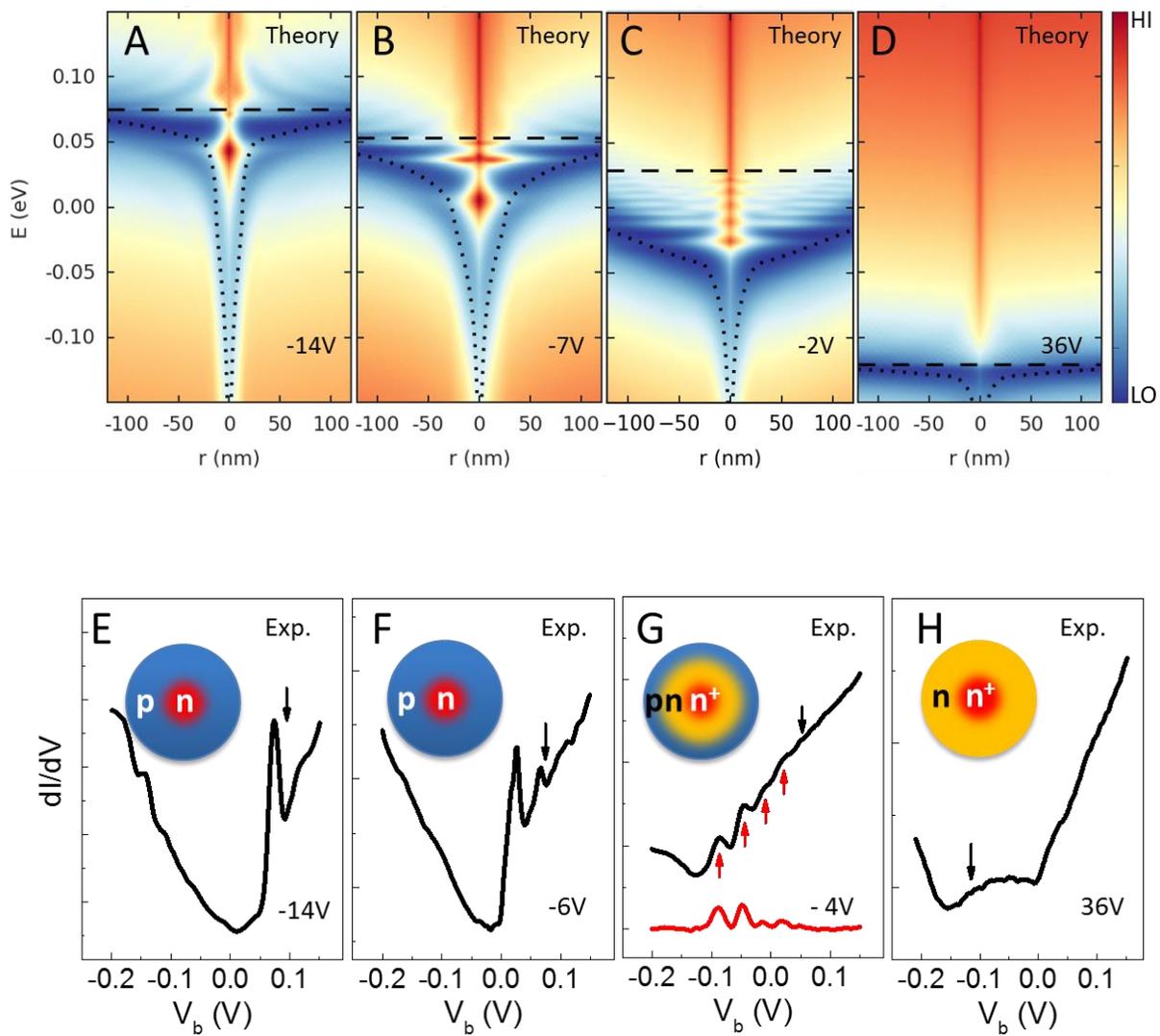

**Figure 3**

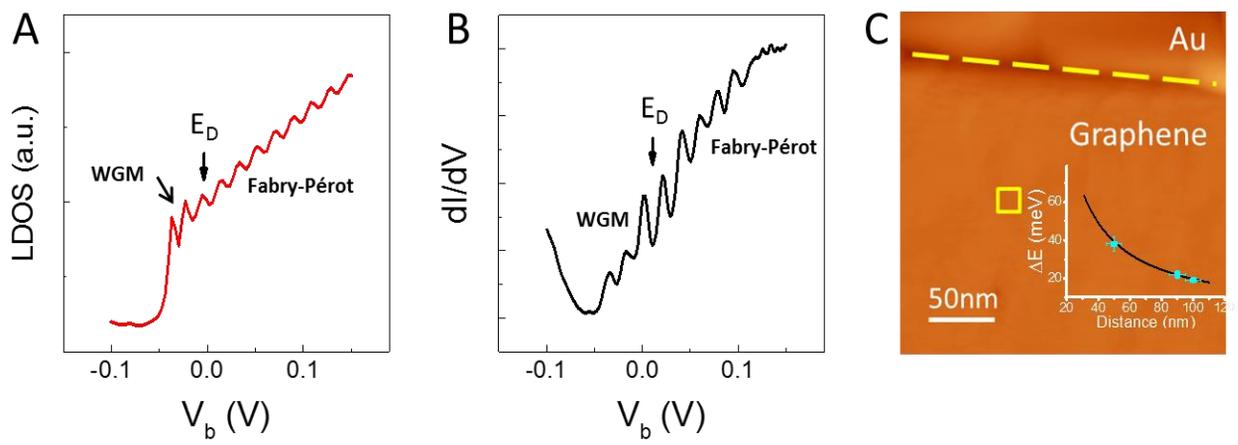

**Figure 4**